\pdfoutput=1 
\documentclass{JINST}

\usepackage{enumitem}
\usepackage{pifont}
\usepackage{setspace}
\usepackage{amssymb,amsmath}

\title{Level-1 pixel based tracking trigger algorithm for LHC upgrade}

\author{Chang-Seong Moon\thanks{Corresponding author, supported by the FP7-PEOPLE-2011-IIF, Contract No. 302103, ``TauKitForNewPhysics''.
}~
and Aurore Savoy-Navarro\\
\llap{}
Laboratoire APC, 
Universit\'{e} Paris Diderot-Paris7/CNRS,\\
10, rue Alice Domon et L\'{e}onie Duquet
75205 Paris Cedex 13, France\\
E-mail: \email{csmoon@cern.ch}}

\abstract{
The Pixel Detector is the innermost detector of the tracking system of the Compact Muon Solenoid 
(CMS) experiment at CERN Large Hadron Collider (LHC). It precisely determines the interaction 
point (primary vertex) of the events and the possible secondary vertexes due to heavy flavours 
($b$ and $c$ quarks); 
it is part of the overall tracking system that allows reconstructing the tracks of the charged particles 
in the events and combined with the magnetic field to measure their momentum.
The pixel detector allows measuring the tracks in the region closest to the interaction point.
The Level-1 (real-time) pixel based tracking trigger is a novel trigger system that is currently 
being studied for the LHC upgrade. An important goal is developing real-time track reconstruction 
algorithms able to cope with very high rates and high flux of data in a very harsh environment.
The pixel detector has an especially crucial role in precisely identifying the primary vertex of the 
rare physics events from the large pile-up (PU) of events. 
The goal of adding the pixel information already at the real-time level of the selection is to 
help reducing the total level-1 trigger rate while keeping an high selection capability.
This is quite an innovative and challenging objective for the experiments upgrade for the High Luminosity 
LHC (HL-LHC).
The special case here addressed is the CMS experiment.
This document describes exercises focusing on the development of a fast pixel track reconstruction 
where the pixel track matches with a Level-1 electron object using a ROOT-based simulation framework.
}


\begin{document}

\section{Introduction}\label{sec:intro}

The LHC experiments ran successfully and provided excellent data at the central collision energy 
of 7 to 8 TeV. 
ATLAS and CMS collaborations have discovered in 2012 a new boson with a mass near 125 GeV and 
properties compatible so far with those predicted for the standard model Higgs boson. 
The discovery of the Higgs boson completes the set of predicted particles by the Standard Model 
(SM) of particle physics. 

The LHC will start operating at the central collision energy of 13 TeV initially in 2015 and progressively 
reaching the design energy of 14 TeV.
This is a new energy frontier.
It will substantially enlarge the mass reach in the search for new particles and will also 
greatly extend the potential to study the properties of the Higgs boson and to explore the overall 
Higgs sector. 
In order to meet the experimental challenges of unprecedented proton-proton luminosity, the CMS 
collaboration will need to improve and upgrade the ability of the apparatus to isolate and 
precisely measure the products of the most interesting collisions. 
Therefore, a key goal of the LHC upgrade will be to maintain the overall physics acceptance 
under the challenging HL-LHC conditions 
(140 PU events in average and an instantaneous luminosity of 5 $\times 10^{34}$ cm$^{-2}$s$^{-1}$). 
CMS must preserve its capability to efficiently trigger events originating from low-mass physics 
processes (e.g. Higgs production at 125 GeV) and for performing precision measurements of low 
to medium transverse momentum ($p_{T}$) physics objects in particular all the leptons 
(electrons, muons and taus), the heavy quarks especially the $b$-quark and consequently the top-quark, 
as well as distinguishing the photons from the electrons.
A series of ongoing feasibility studies are conducted within the CMS experiment; 
they show the benefits of having the pixel information included in the Level-1 (L1) track trigger. 
The next crucial challenge is of course to show the real possibility to include it in a realistic scenario taking into account the constraints in bandwidth and latency at the L1 trigger in the LHC experiments (ATLAS and CMS).

Section~\ref{sec:pixelupgrade} will briefly describe the CMS pixel detector and its sequential upgrades in Phase 1 and Phase 2. We introduce in Section~\ref{sec:pixtrk}, as an example, a L1 pixel tracking algorithm, called PiXTRK. It is a first attempt to achieve a real-time track reconstruction using the pixel clusters within a region seeded by the L1 electromagnetic (EM) calorimeter trigger tower.
Section~\ref{sec:result} briefly reports the main results obtained by the feasibility studies on the potential of L1 pixel trigger in CMS.
Section~\ref{sec:exercise} describes in more details the exercises given to the students in the international Summer School on "INtelligent Signal Processing for FrontIEr Research and Industry" held in Paris~\cite{infieri}.

\section{The upgrade of the Pixel detector from Phase 1 to Phase 2}\label{sec:pixelupgrade}

\begin{figure}[tbp] 
\centering
\includegraphics[width=0.70\textwidth]{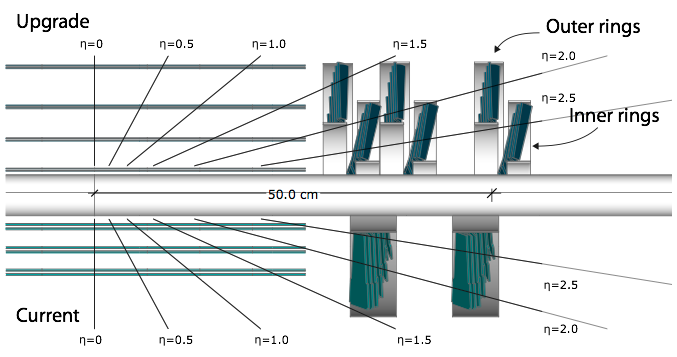} \\
\includegraphics[width=0.80\textwidth]{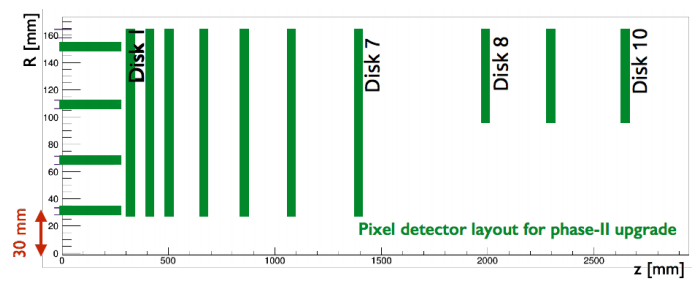}
\caption{The top picture actually shows the Phase 1 upgrade above the beam pipe and the existing detector below beam pipe as a comparison, and the bottom layout shows the Phase 2 upgrade of the pixel detector}
\label{fig:phase12}
\end{figure}

The present CMS pixel detector consists of three barrel layers with two endcap disks that covers 
a pseudorapidity range $|\eta|<$ 2.5, matching the acceptance of the central tracker~\cite{cms_lhc}.
A first upgrade for Phase 1 (around 2018) of the current pixel detector will add a fourth barrel layer, a third disk on each endcap sides, as well as a new Front End ASIC including signal digitization, and consequently a new readout chain with higher performance in speed and bandwidth.
The main motivation for the Phase 1 upgrade of the pixel detector is to maintain the high efficiencies 
and low fake rates and to minimize the data loss due to latencies and limited buffering in higher luminosity running conditions~\cite{phase1pixel}.
In the Phase 2 upgrade for HL-LHC, the CMS pixel detector will have an extended pseudorapidity 
coverage up to $|\eta| < 4$ as shown in figure~\ref{fig:phase12}~\cite{phase2pixel}.
It will be made of new pixel sensors (the sensor technology is being selected) and of smaller size pixels 
(higher granularity).
A new Front End and readout system is currently developed within the R\&D (RD53) Collaboration~\cite{rd_coll} for the HL-LHC era (2023-2025).

\section{The L1 pixel based Electron trigger algorithm}\label{sec:pixtrk}

\begin{figure}[tbp] 
\centering
\includegraphics[width=0.9\textwidth]{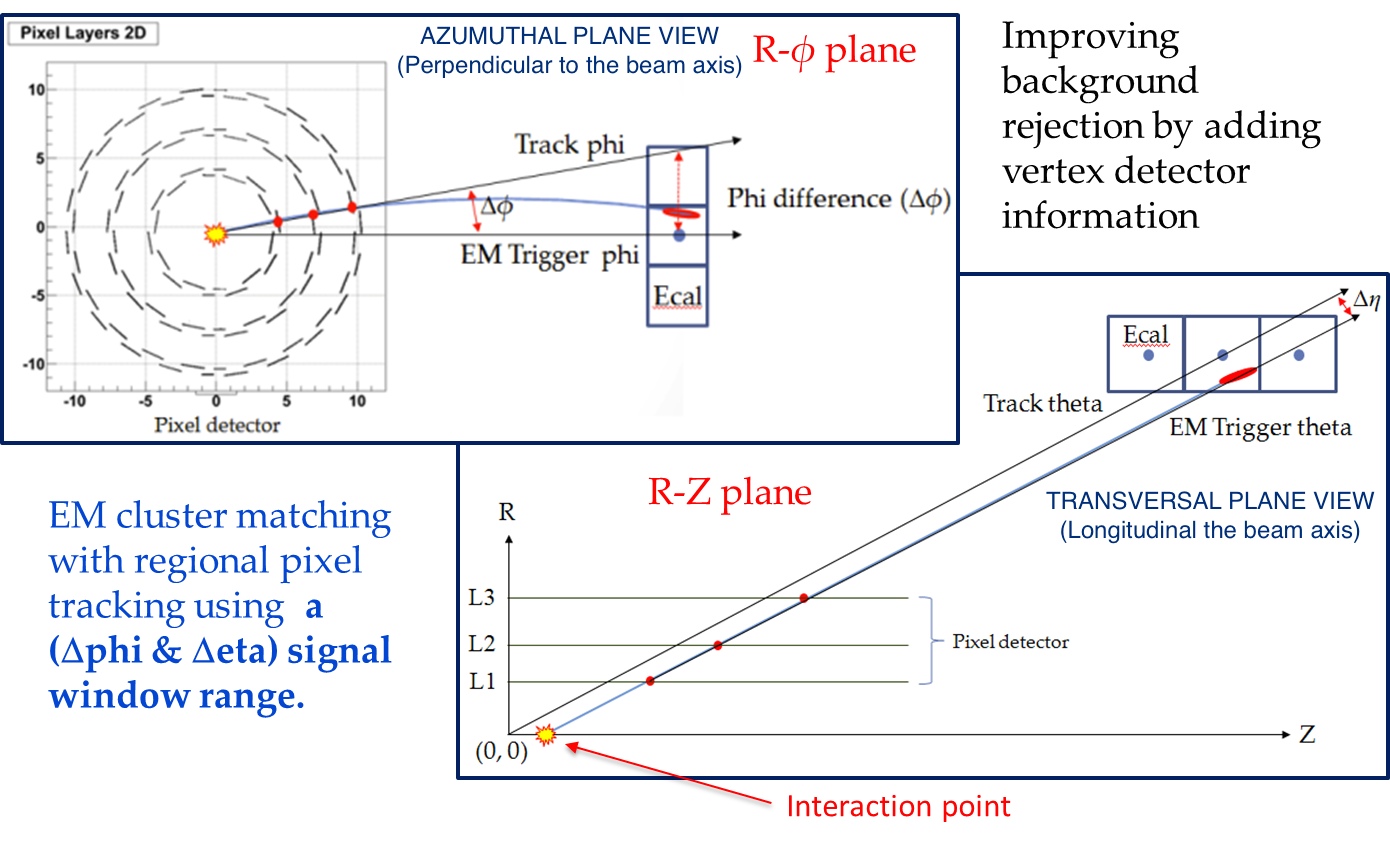}
\caption{Schema summarizing the PiXTRK pattern recognition strategy in the azimuthal and transversal plane views for the current pixel barrel layout}
\label{fig:pixtrk}
\end{figure}

\begin{figure}[tbp] 
\centering
\includegraphics[width=0.5\textwidth]{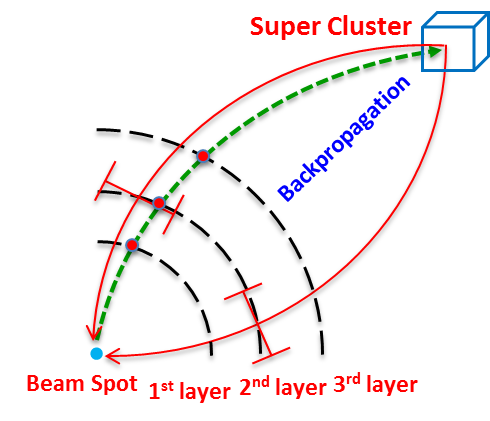}
\caption{Schema describing the Pixel matching with the Super Cluster in HLT}
\label{fig:pixtrkHLT}
\end{figure}

This series of exercises concentrates on the application of the pixel detector to the electron trigger.
The PiXTRK algorithm we are using here and that has been developed for ongoing feasibility studies in CMS for HL-LHC, performs the {\it pattern recognition}. This is the first step for the track reconstruction. To do so, PiXTRK transfers to Level-1 what is actually achieved in the first stage of the current CMS High Level Trigger (HLT), namely the matching of pixel hits with the EM cluster \cite{HLT}.
The pattern recognition proceeds by first defining $\Delta \phi$ windows ($i.e.$ in the transverse plane to the beam axis), seeded by the EM cluster, and going backward to the beam spot (figure \ref{fig:pixtrkHLT}).
The limit conditions imposed on all signal windows vary as a function of EM $E_T$.
The signal window are set within 3$\sigma$ standard deviation. It defines the area in the corresponding plot containing 99.9\% of the hits.
Furthermore PiXTRK distinguishes between positive and negative charged particles, here electrons and positrons.
But unlike the HLT case because of bandwidth limitation at Level-1, it is not possible to use individual pixel hits: we thus consider instead pixel clusters. It is not possible to benefit from the refined EM cluster calorimeter reconstruction here, as only the EM clusters defined by the L1 calorimeter trigger are available.
The pattern recognition proceeds then in 2 stages which are described in a bit more detail in subsections~\ref{sec:pixEMmatching} and \ref{sec:standalonePix}.

\subsection{Pattern recognition seeded by the L1 EM cluster}\label{sec:pixEMmatching}

It follows two steps:

\subsubsection{Pattern recognition in the $R-\phi$ transverse plane}\label{sec:PR1}

\begin{figure}[tbp] 
\centering 
  \begin{tabular}[t]{cc}
\includegraphics[width=0.40\textwidth]{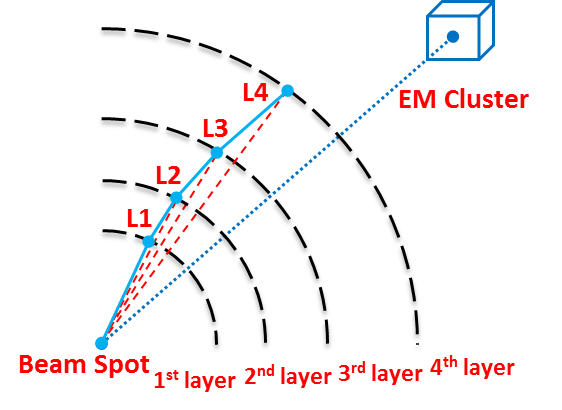} &
\includegraphics[width=0.40\textwidth]{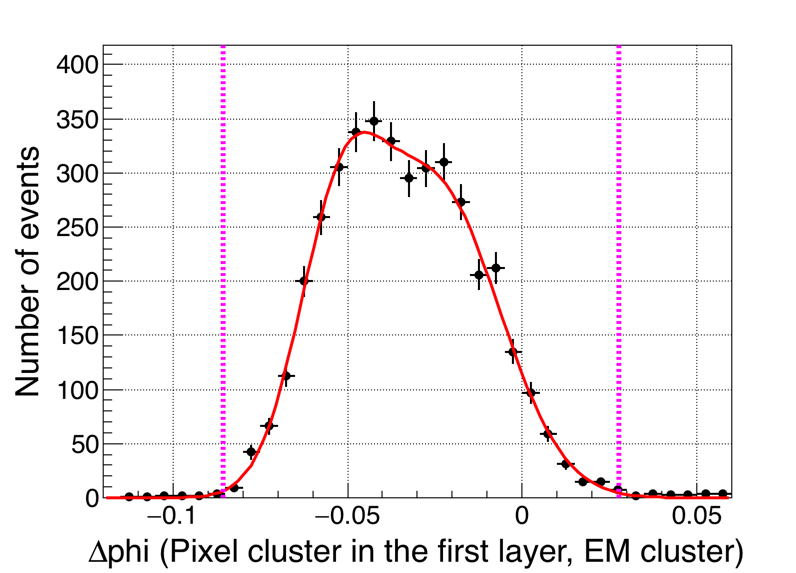} \\
(a) & (b) \\
  \end{tabular}
\caption{Pixel track matching with EM cluster in the azimuthal plane: (a) for each pixel layer $L_i$ one looks to the $\phi$ azimuthal angle between the pixel segment $(BS, L_i)$ joining the beam spot (BS) and the relevant pixel cluster in the $L_i$ layer and the segment $(BS, EM)$ joining the beam spot and the EM cluster (EM);
(b) shows as an example the $\Delta \phi = \phi(BS, L_1) - \phi(BS, EM)$ plot for single electron MC data with momentum 20 to 21 GeV, no PU, and setting the signal selection boundary at 3$\sigma$ standard deviation.}
\label{fig:pixEM1}
\end{figure}

Firstly the region of interest (RoI) is defined, in the transverse $R-\phi$ plane, by the L1 EM cluster linked to the beam spot (BS). The selected pixel clusters in each layer are those which are in the $\Delta \phi$ window (figure \ref{fig:pixEM1}) defined as:
\begin{equation}\label{eq:1}
\Delta \phi = \phi(BS, L_i) - \phi(BS, EM) < 0.1 
\end{equation}
Where $(BS, L_i)$ is the pixel segment joining the beam spot with the relevant pixel cluster in the corresponding $L_i$ layer. And the segment of $(BS, EM)$ joins the beam spot with the L1 EM cluster.
In this region are selected the pixel clusters which, in each layer ($L_i$, $i$=1,$\ldots$,4) are included  in a $\Delta \phi$ window defined here by $\Delta \phi <$ 0.1, and considering both the cases of electrons and positrons.
In the case of more than one cluster satisfying the equation \ref{eq:1}, all the combinations corresponding to all the possible clusters in this region of interest are considered.
The pattern recognition procedure is further refined with the following two steps.

\subsubsection{Refined Pattern recognition seeded by the EM cluster}

\begin{figure}[tbp] 
\centering 
  \begin{tabular}[t]{cc}
\includegraphics[width=0.40\textwidth]{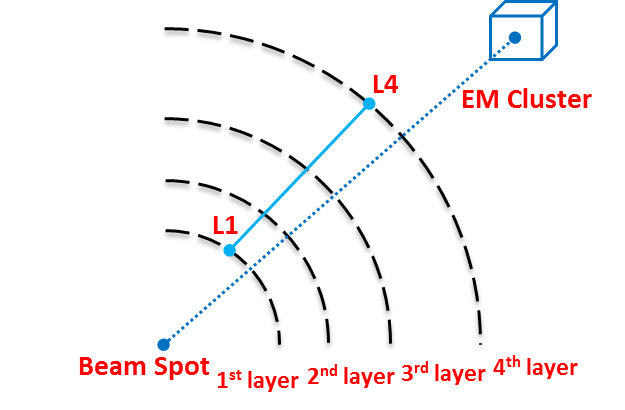} &
\includegraphics[width=0.40\textwidth]{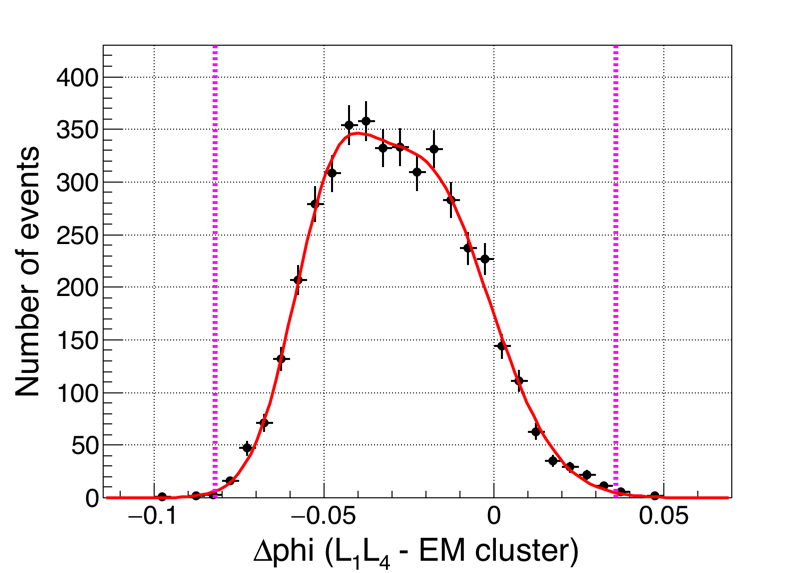} \\
(a) & (b) \\
\includegraphics[width=0.40\textwidth]{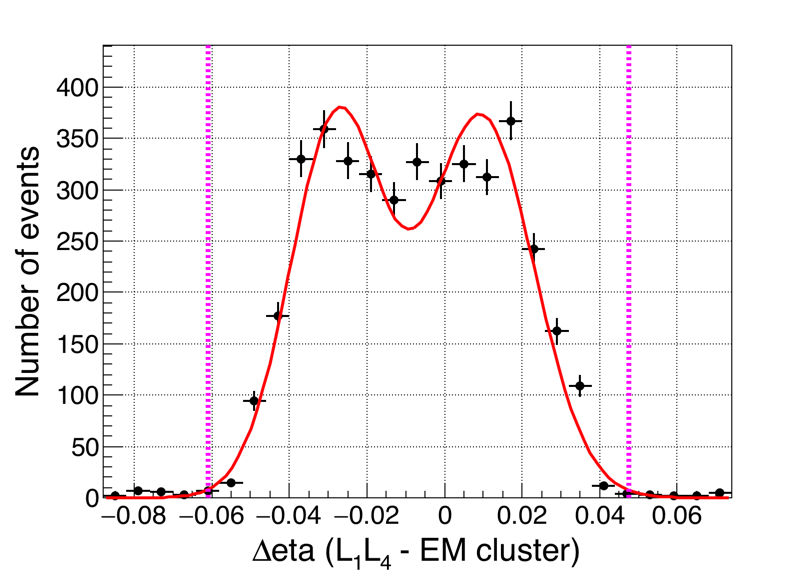} &
\includegraphics[width=0.40\textwidth]{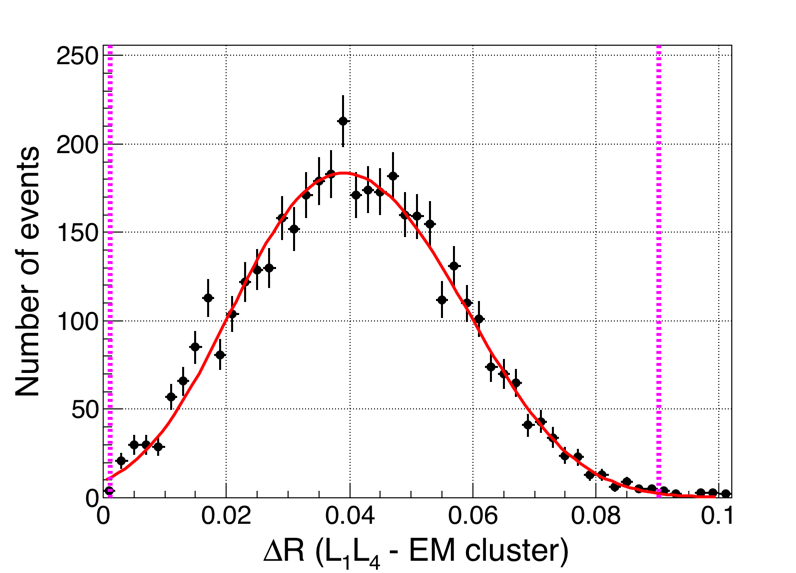} \\
(c) & (d) \\
  \end{tabular}
\caption{Pixel track matching with EM cluster in $\Delta \eta$, $\Delta \phi$ and $\Delta R$, for each combination of a pair of pixel layers; (a) shows, as an example, the combination of a pair corresponding to $L_1$ and $L_4$ layers, with ($L_1, L_4)$ segment matching in $\phi$ with $(BS, EM)$ segment; (b) corresponding $\Delta \phi = \phi(L_1, L_4) - \phi(BS, EM)$ distribution; (c) corresponding $\Delta \eta = \eta(L_1, L_4) - \eta(BS, EM)$ distribution (the double peak structure is due to the 2x2 Level-1 EM tower granularity); (d) corresponding $\Delta R = \sqrt{(\Delta\eta)^2 + (\Delta\phi)^2}$; All the distributions include the signal selection boundary at 3$\sigma$ and correspond to EM transverse energy ($E_T$) range from 20 to 21 GeV.}
\label{fig:pixEM2}
\end{figure}

This step involves in determining the $\Delta \eta$, $\Delta \phi$ and $\Delta R$ signal windows, as a function of EM $E_T$, for the set of the pixel clusters selected by the condition defined by the equation \ref{eq:1}, for each layer. Each combination of a pair of pixel layers is considered to form all the possible $(L_i, L_j)$ track segments which correspond to each pixel cluster retained in the selection defined by the condition (\ref{eq:1}). It compares the matching of each of these $(L_i, L_j)$ segments with the segment joining the beam spot with the EM cluster in the ($\eta$, $\phi$) coordinates by defining:
\begin{equation}
\begin{split}
\Delta \eta = \eta(L_i, L_j) - \eta(BS, EM) < 3\sigma \\
\Delta \phi = \phi(L_i, L_j) - \phi(BS, EM) < 3\sigma  \\
\rm{and}~\Delta R = \sqrt{(\Delta\eta)^2 + (\Delta\phi)^2} < 3\sigma 
\end{split}
\end{equation}
Where $i$, $j$ = 1,$\ldots$,4 and $i \neq j$.
The pixel cluster in each layer is then selected only if it passes the 3$\sigma$ signal window requirements in $\Delta \phi$, $\Delta \eta$ and $\Delta R$ as shown in figure \ref{fig:pixEM2}.

\subsection{The standalone pattern recognition}\label{sec:standalonePix}

\begin{figure}[tbp] 
\centering
  \begin{tabular}[t]{ccc}
  & & \\
\includegraphics[width=0.31\textwidth]{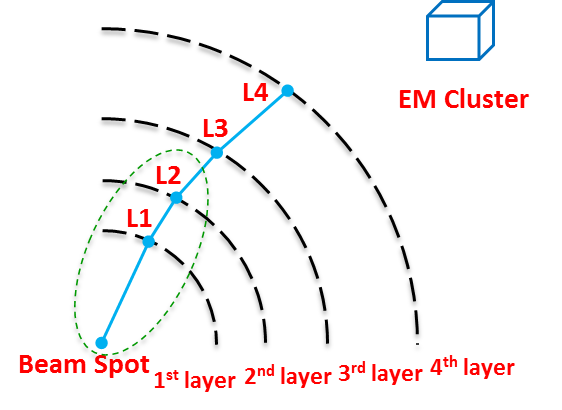} &
\includegraphics[width=0.31\textwidth]{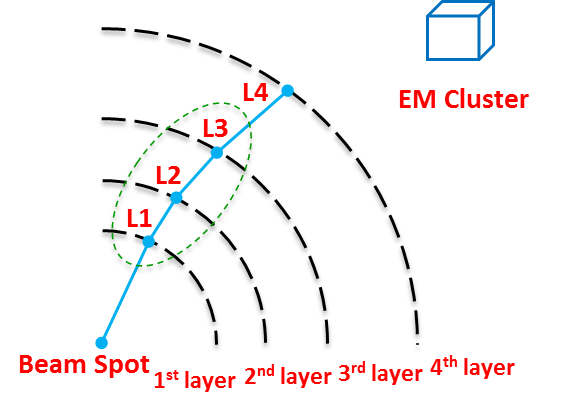} &
\includegraphics[width=0.31\textwidth]{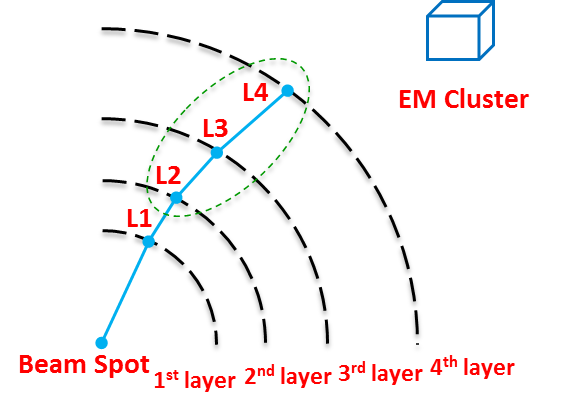} \\
 & (a) & \\
\includegraphics[width=0.31\textwidth]{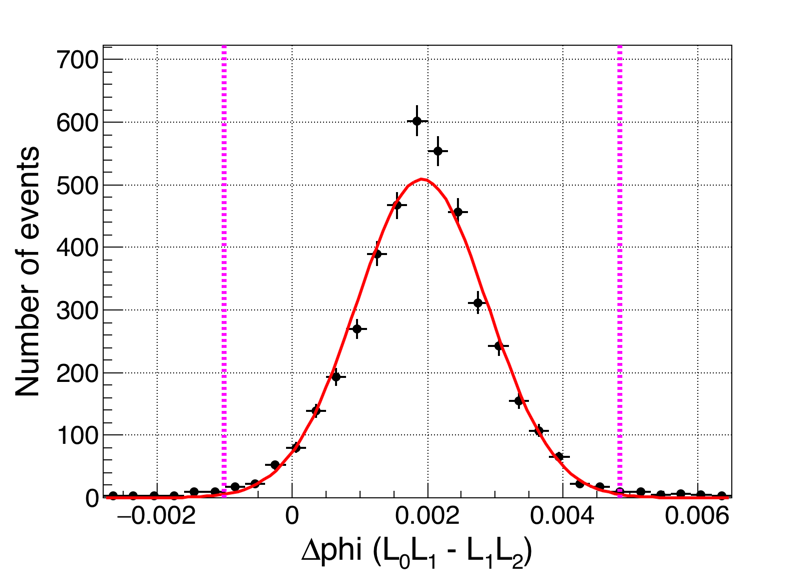} &
\includegraphics[width=0.31\textwidth]{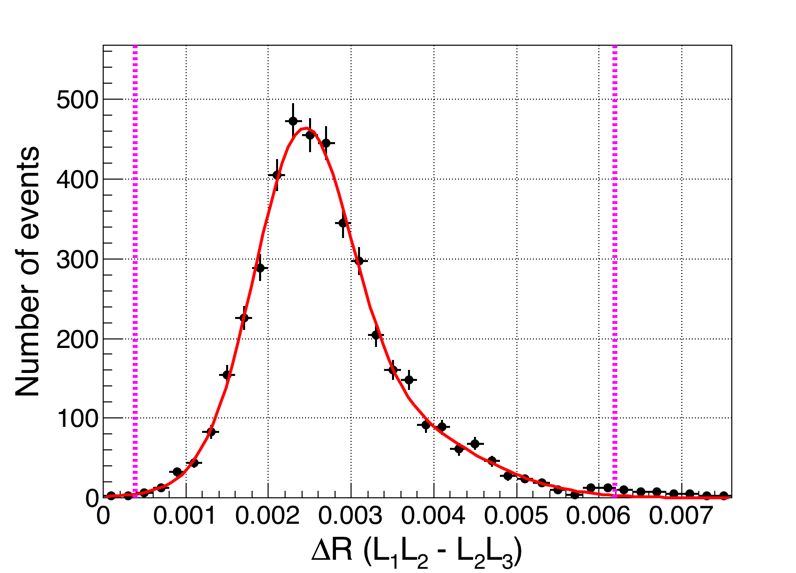} &
\includegraphics[width=0.31\textwidth]{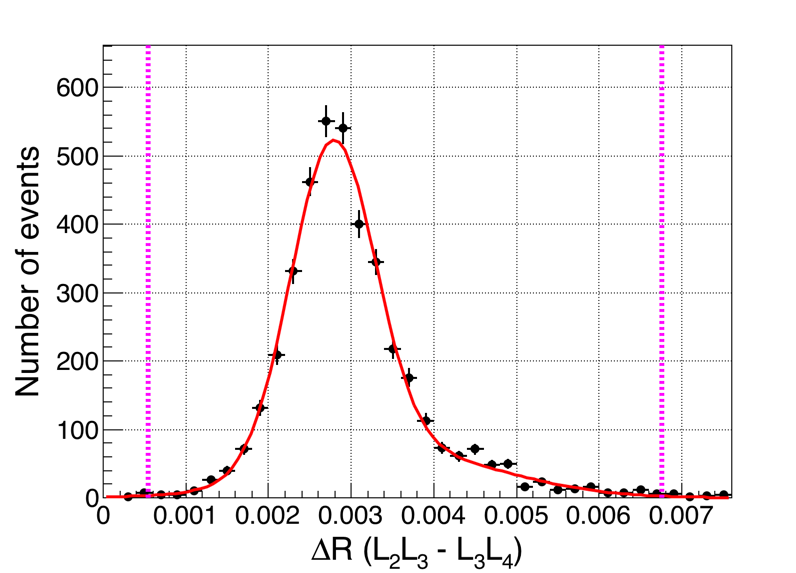} \\
 & (b) & \\
  \end{tabular}
\caption{Standalone pattern recognition with pixel clusters: (a) top views shows all possible three aligned clusters cases including also BS.
(b) Left plot shows the corresponding $\Delta \phi = \phi (L_0, L_1) - \phi (L_1, L_2)$ to the top left schema.
Middle plot shows the $\Delta R$ corresponding to the top middle schema and right plot shows the $\Delta R$ corresponding to the top right schema. All the distributions include the three sigma boundary and correspond to EM transverse energy ($E_T$) range from 20 to 21 GeV.
}
\label{fig:standpixtrk}
\end{figure}

This second step aims to further reduce fake combinations.
To do so, we then consider all the possible 3-layer combinations, with surviving 2-layer vector selection in each of the 4 layers including also the beam spot.
The $\Delta \phi$, $\Delta \eta$ and $\Delta R$ signal windows (figure \ref{fig:standpixtrk}), as a function of EM $E_T$, are now defined by:
\begin{equation}
\begin{split}
\Delta \eta = \eta (L_i, L_j) - \eta (L_j, L_k) < 3\sigma \\
\Delta \phi = \phi (L_i, L_j) - \phi (L_j, L_k) < 3\sigma \\
\rm{and}~\Delta R = \sqrt{(\Delta\eta)^2 + (\Delta\phi)^2} < 3\sigma
\end{split}
\end{equation}
Where $i, j, k$ = 0,$\ldots$,4 with $L_0 = BS$ and $i \neq j \neq k$.
The pixel cluster must satisfy all the signal windows requirements within 3 standard deviations.
The complete procedure progressively achieves good pattern recognition through the series of $\Delta \eta$, $\Delta \phi$ and $\Delta R$ window cuts. The obtained track hit cluster collections can then be used for the track fitting as the next stage of track reconstruction which is beyond the scope of this exercise.

\section{Preliminary outcomes from feasibility studies on a L1 pixel based electron trigger}\label{sec:result}

To achieve this exercise, two samples of fully simulated MC data corresponding to the new CMS tracking system for the Phase 2 for HL-LHC and two cases, namely: i) single electrons with superimposed minimum bias data for simulating the PU and ii) minimum bias data for evaluating fake electrons. 
Applying then the PiXTRK algorithm to the case of a L1 trigger on electrons for HL-LHC with an average mean value ($\mu$) of 140 PU, overall efficiency of 93\% in the central barrel and a rate reduction factor for L1 electron trigger of 8 for a L1 EM $E_T$ threshold of 20 GeV are obtained.
This is a first very encouraging first indication.
The improved L1 electron trigger rate reduction by requiring a matched L1 pixel trigger, allows a significantly lower threshold for the L1 EM $E_T$ trigger even at an average PU of 140. This study implies a possible electron trigger track reconstruction using just pixel clusters seeded by the EM trigger tower (see section~\ref{sec:pixtrk}).
A series of other feasibility studies in CMS are underway for the extension in $\eta$ of this trigger and for the use for $b$-tagging in the first level of trigger (using the L1 outer track as a seed).

\section{EXERCISES}\label{sec:exercise}

\begin{figure}[tbp] 
\centering
  \begin{tabular}[t]{cc}
\includegraphics[width=0.31\textwidth]{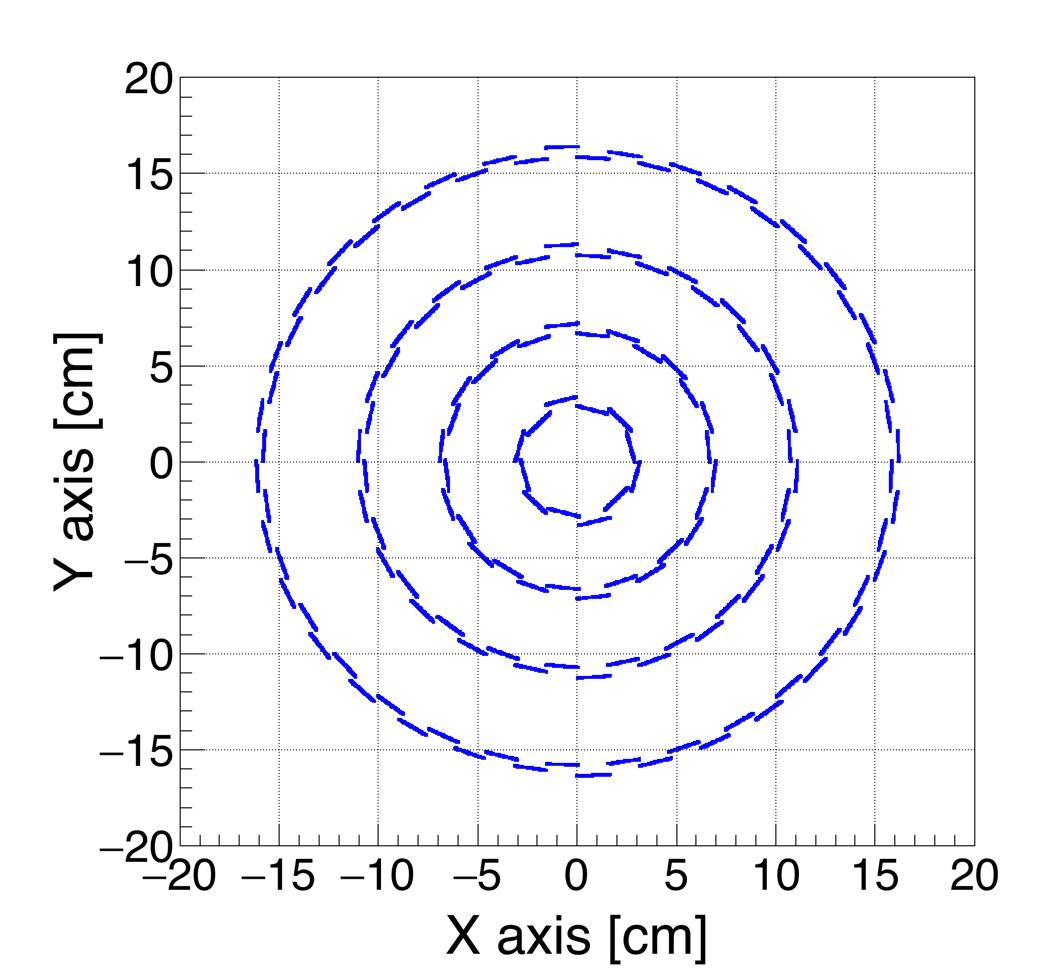} &
\includegraphics[width=0.31\textwidth]{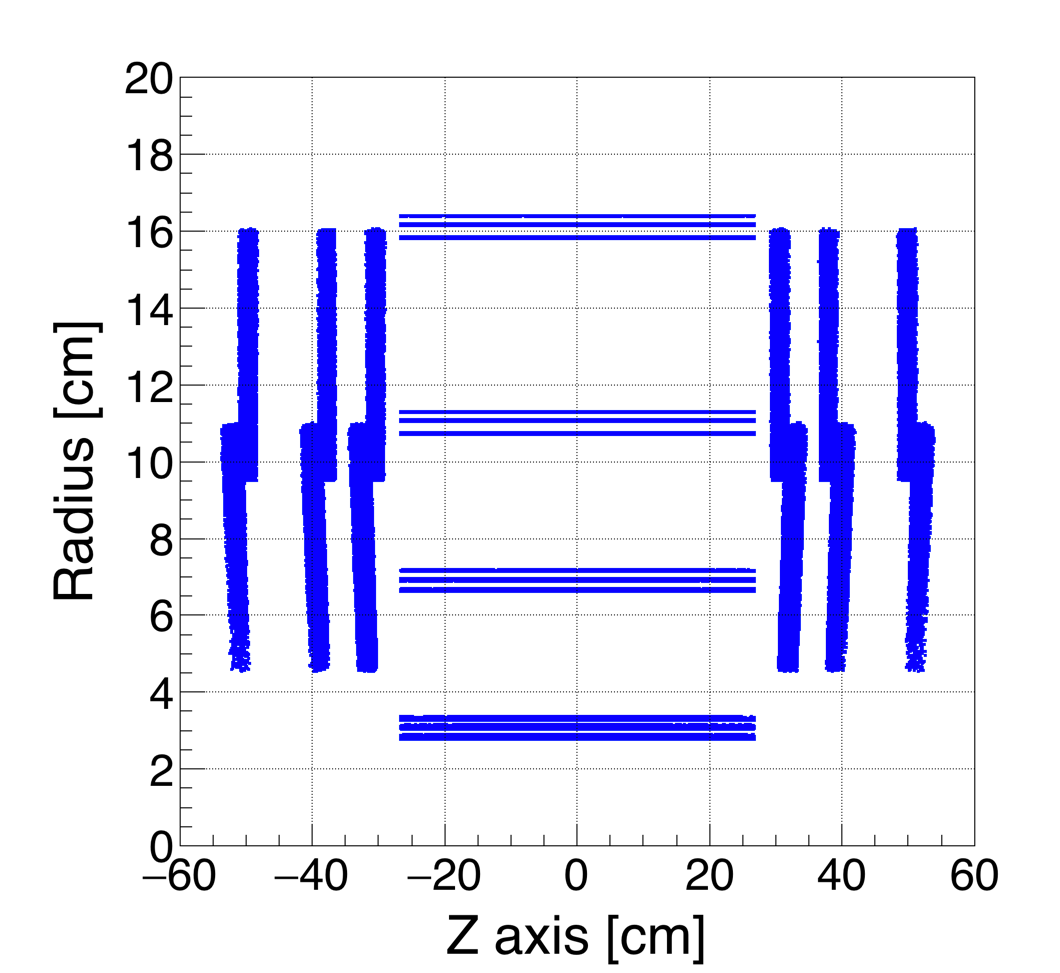} \\
(a) & (b) \\
  \end{tabular}
  \begin{tabular}[t]{ccc}
\includegraphics[width=0.31\textwidth]{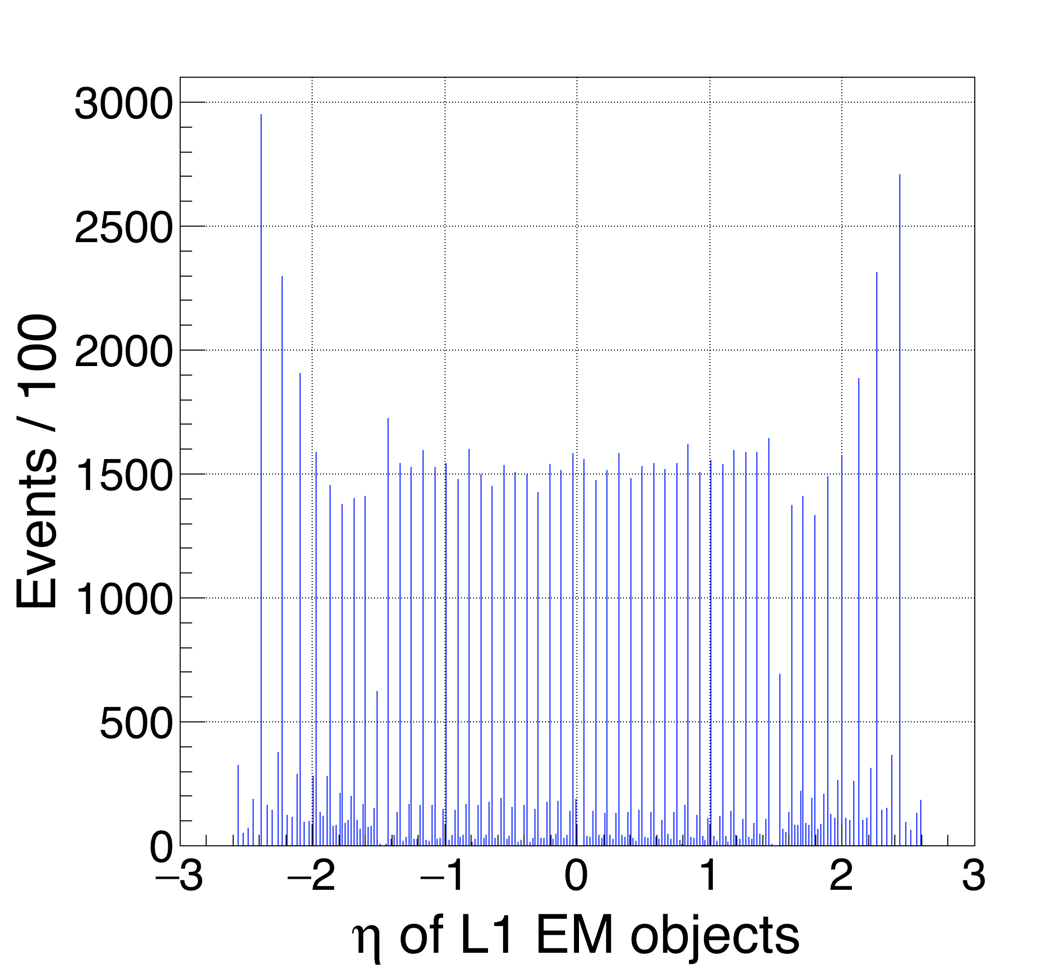} &
\includegraphics[width=0.31\textwidth]{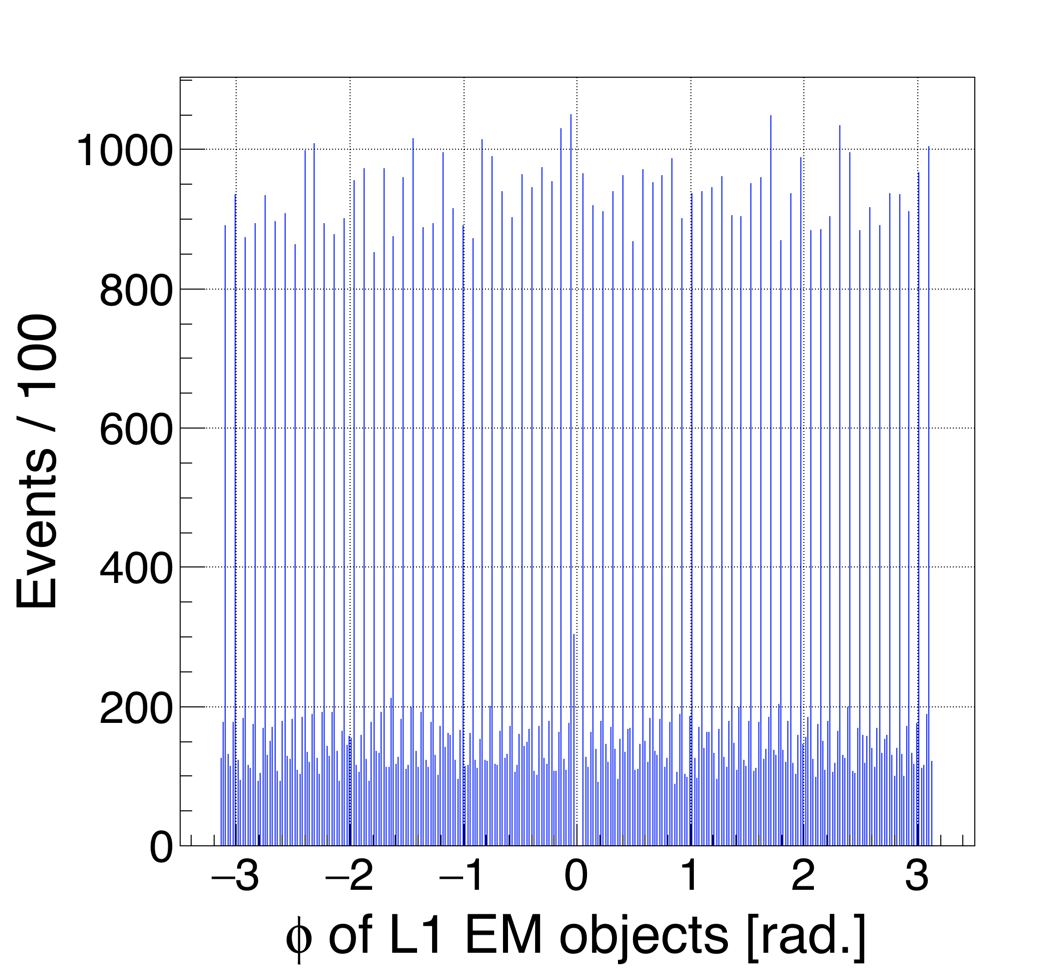} &
\includegraphics[width=0.31\textwidth]{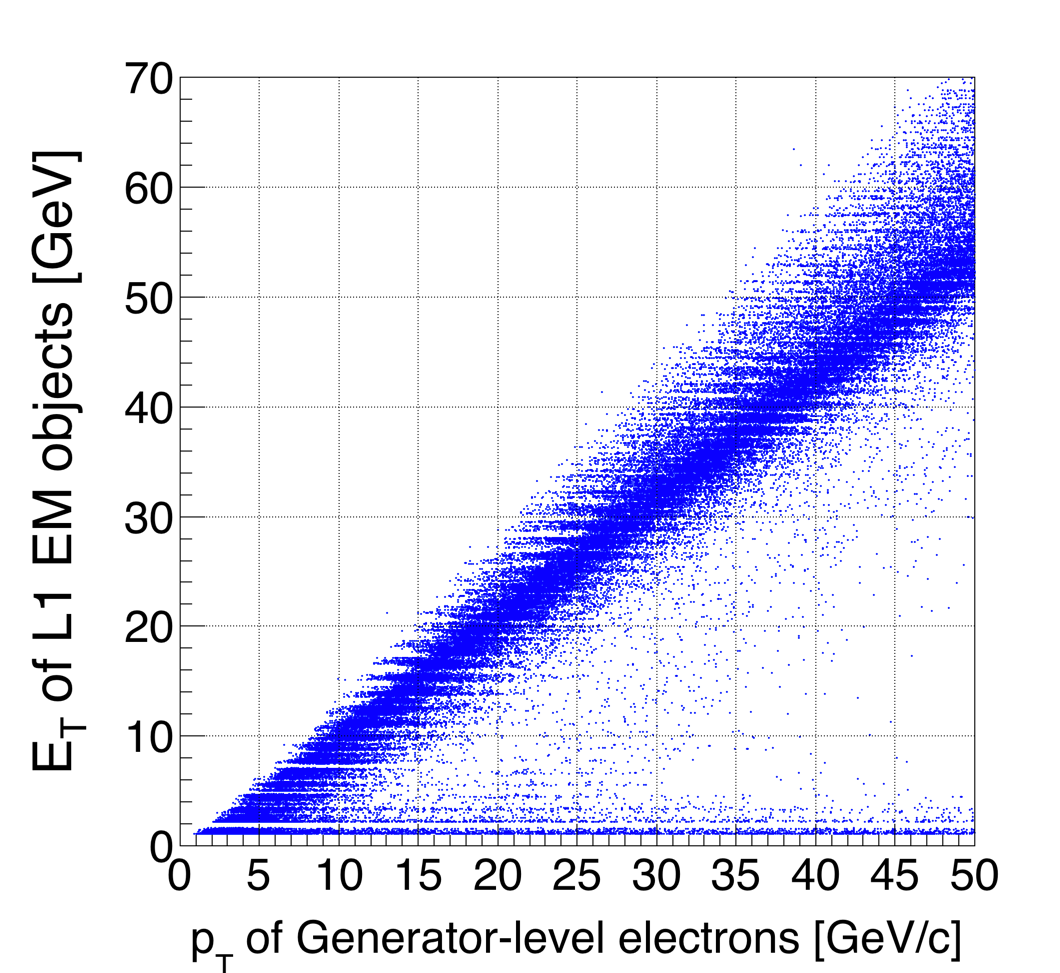} \\
(c) & (d) & (e) \\
  \end{tabular}
\caption{EXERCISE 1 results: Exercise 1-1 (a): view of pixel layers in X-Y plane; Exercise 1-2 (b): view of pixel layers and disks in R-Z plane; Exercise 1-3 (c): $\eta$ distribution of L1 EM clusters; Exercise 1-4 (d):  $\phi$ distribution of L1 EM clusters; Exercise 1-5 (e): 2-dimensional distribution of gen-level electron $E_T$ vs. L1 EM $E_T$}
\label{fig:ex1}
\end{figure}

The exercises for introducing the students to a L1 pixel-based tracking trigger algorithm are made of the following series of exercises divided into four sections.
\begin{itemize}
\raggedright
  \item[(1)] 
In the exercise 1-1, the students are introduced to the ROOT-based software framework and the CMS pixel data format. In exercises 1-2 \& 1-3 (figure\ref{fig:ex1} (a) and (b)), the students learn the CMS pixel detector geometry.
In exercise 1-4, the students get to also understand the granularity of the EM calorimeter and the size in elementary cells that defines the L1 EM trigger tower (figure\ref{fig:ex1} (c) and (d)).
The dimension of the L1 EM tower is instrumental for defining the "region of interest" as 
the L1 trigger tower size defines the dimension of the seed in this case.
The last step in the first exercise is to examine the correlation between the MC generator-level truth for the electron and the L1 EM cluster to verify the electron identification. Figure \ref{fig:ex1}(e) shows the measured transverse energy agrees well with MC generator-level truth within expected resolution.

\renewcommand{\arraystretch}{1.3}
\renewcommand{\tabcolsep}{1.0mm}
\begin{tabular}{rlp{11.5cm}}
- & Exercise 1-1 : & Understanding the Ntuple data structure in the ROOT framework \\
- & Exercise 1-2 : & Drawing the pixel geometry in the X-Y plane using pixel clusters on each pixel layer (figure \ref{fig:ex1} (a))  \\
- & Exercise 1-3 : & Drawing the pixel geometry in the R-Z plane using pixel clusters on each pixel layer and each    disk (figure \ref{fig:ex1} (b))\\
- & Exercise 1-4 : & Plotting separately the L1 EM cluster $\eta$ and $\phi$ distributions (figure \ref{fig:ex1} (c) and (d))\\
- & Exercise 1-5 : & Plotting the 2-dimensional plot between the generation-level electron transverse momentum ($p_{T}$) vs. the L1 EM $E_T$ (figure \ref{fig:ex1} (e)). \\
\end{tabular}
\end{itemize}

\begin{figure}[tbp] 
\centering
  \begin{tabular}[t]{ccc}
\includegraphics[width=0.31\textwidth]{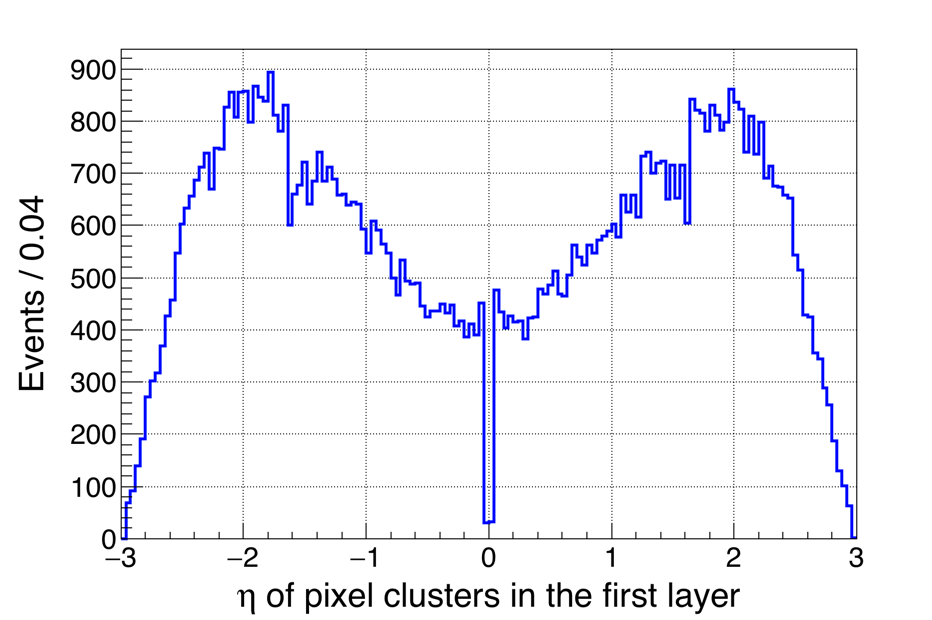} &
\includegraphics[width=0.31\textwidth]{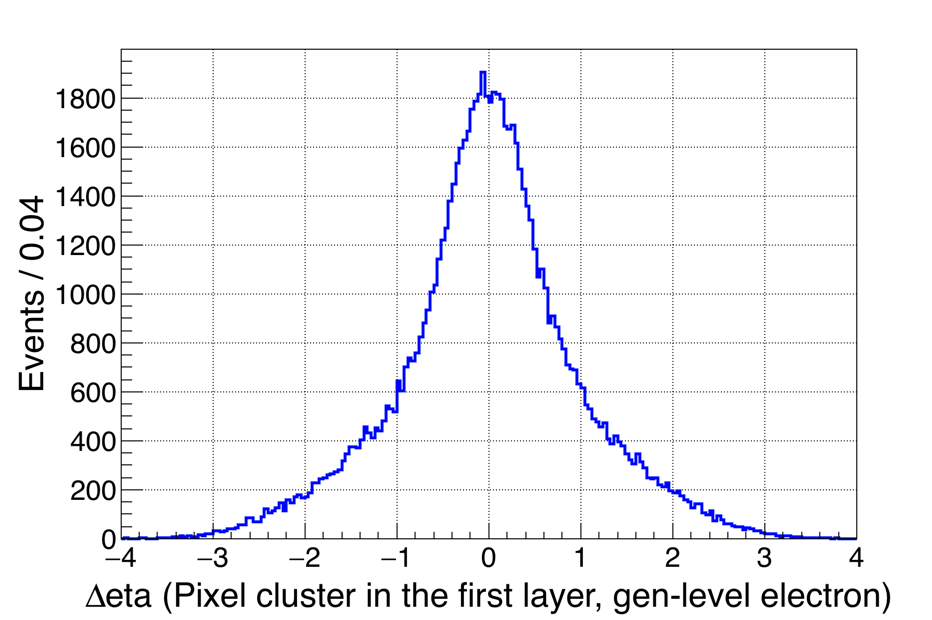} &
\includegraphics[width=0.31\textwidth]{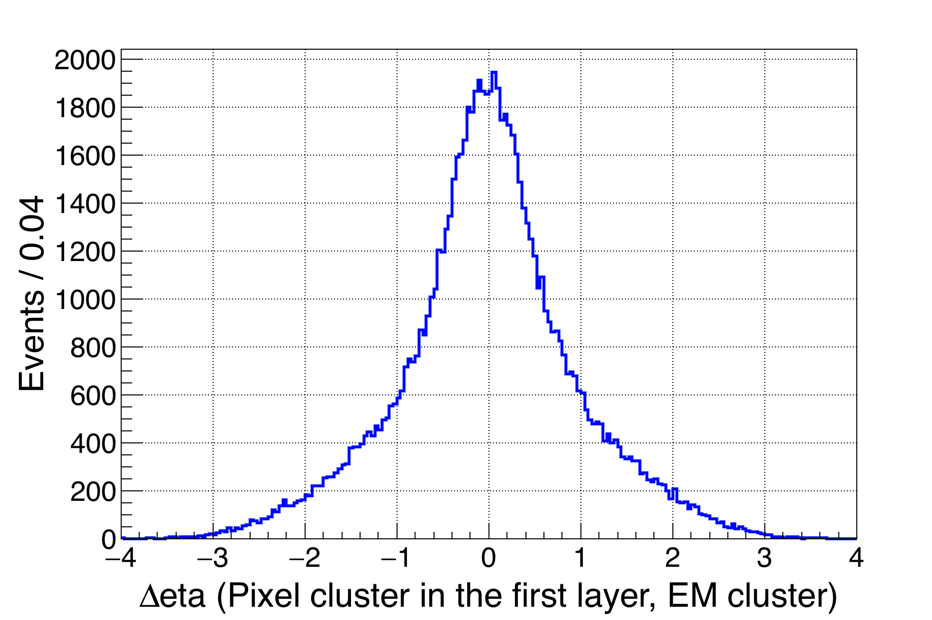} \\
(a) & (c) & (e) \\
\includegraphics[width=0.31\textwidth]{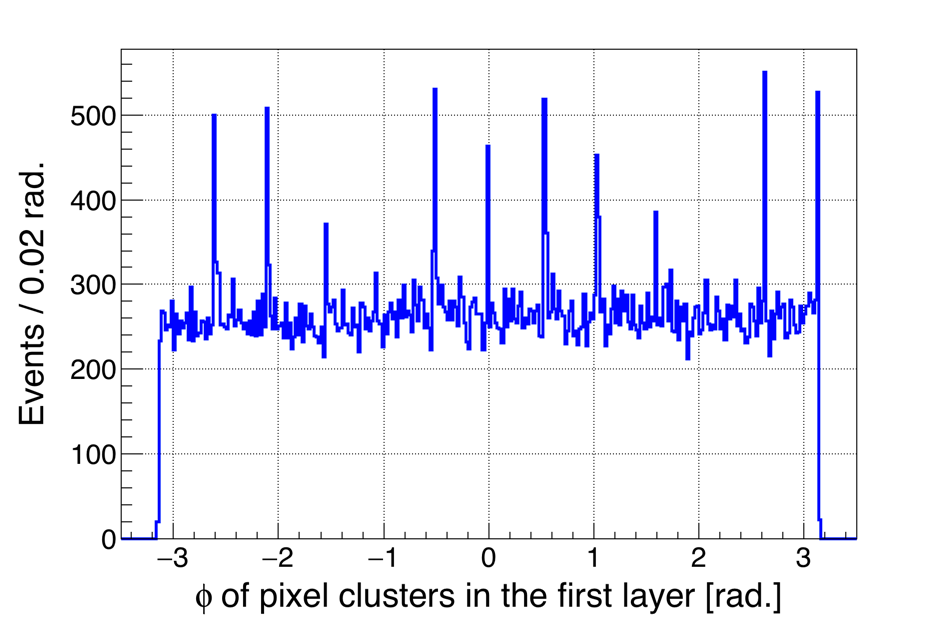} &
\includegraphics[width=0.31\textwidth]{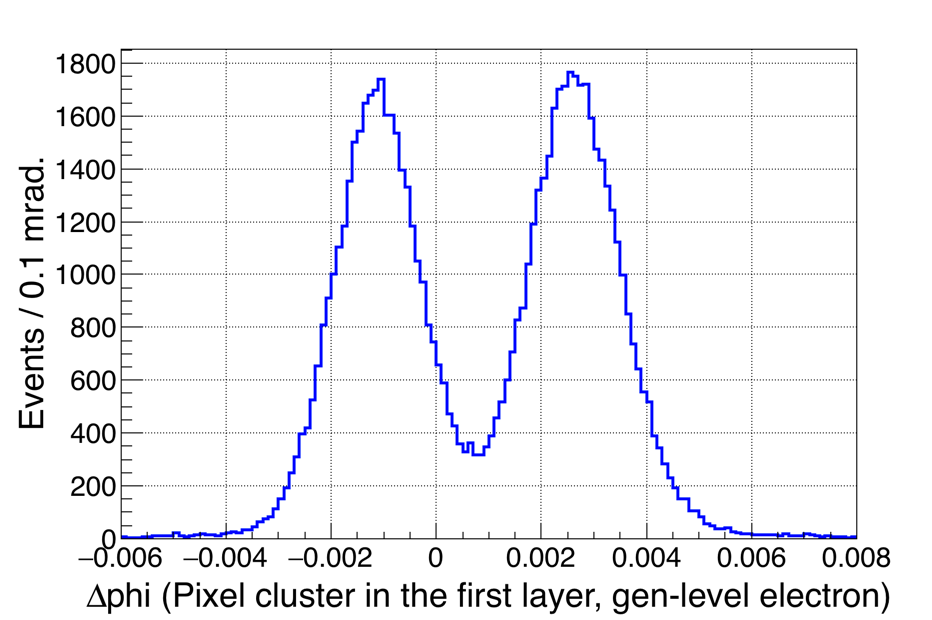} &
\includegraphics[width=0.31\textwidth]{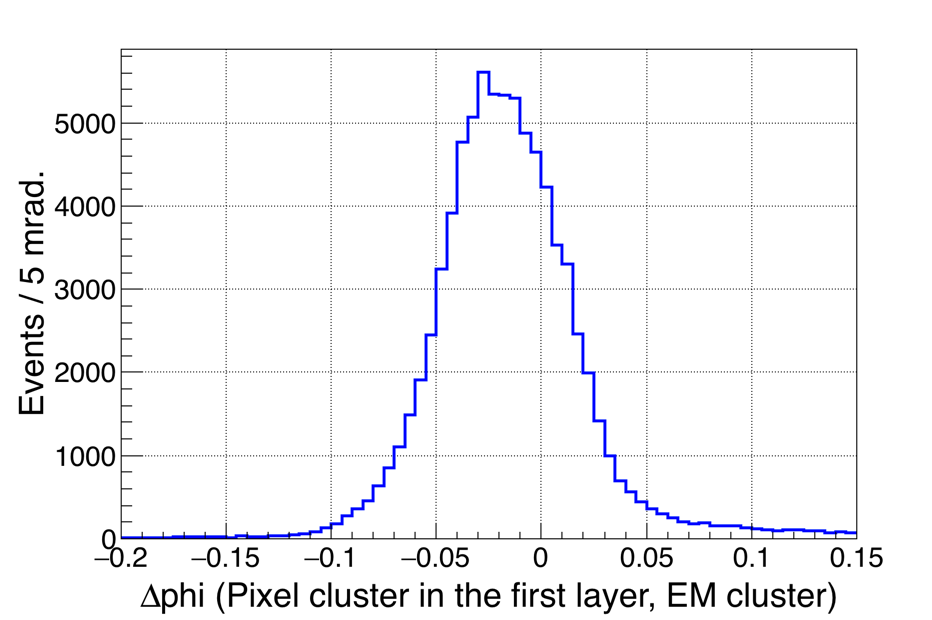} \\
(b) & (d) & (f) \\
  \end{tabular}
\caption{EXERCISE 2 results: Exercise 2-1: $\eta$ (a) and $\phi$ (b) distributions of pixel clusters in the first pixel layer; Exercise 2-2: $\Delta\eta$ (c) and $\Delta\phi$ (d) distributions between pixel clusters in the first pixel layer and gen-level electron; Exercise 2-3: $\Delta\eta$ (e) and $\Delta\phi$ (f) distributions between pixel clusters in the first pixel layer and L1 EM cluster}
\label{fig:ex2}
\end{figure}

\begin{itemize}
\raggedright
  \item[(2)] The purpose of the second exercise is to match the track reconstructed from pixel clusters using the generator-level electron and the L1 EM cluster.
In the exercise 2-1, the students have to simply calculate the $\eta$ and $\phi$ angles in cylindrical coordinates from the pixel clusters defined in cartesian coordinates (figures \ref{fig:ex2} (a) and (b)).
One should note the spikes in figure \ref{fig:ex2} (b); they indicate the adjacent sensor overlap. If BS would be different from the origin (0,0) as this is the case in reality, these spikes would be instead replaced by deeps that correspond to gaps in coverage.

Using the previously obtained $\eta$ and $\phi$ coordinates, the $\Delta\eta$ and $\Delta\phi$ distributions defined as the difference between the $\eta$ (respectively $\phi$) coordinate of the pixel cluster on each layer and the $\eta$ (respectively $\phi$) of the generator-level electron or the $\eta$ (respectively $\phi$) of the L1 EM cluster are plotted in the exercises 2-2 \& 2-3 (figures \ref{fig:ex2} (c) and (d)). The double peak in figure \ref{fig:ex2} (d) is due to pixel resolution ($\sim$180 $\mu m$).

This step is for understanding how precisely one can determine the $\Delta\eta$ and $\Delta\phi$ signal windows using the pixel detector and the L1 EM calorimeter (figures \ref{fig:ex2} (e) and (f)). The shift in the center of figure \ref{fig:ex2} (f) reflects the difference in curvature of the electron track as measured by the pixel layer 1 (3 cm distance from BS) and at the EM cluster level (1.4m distance from BS). Note that we use here a single electron gun data sample with average $p_T$ of 30 GeV.

\renewcommand{\arraystretch}{1.3}
\renewcommand{\tabcolsep}{1.0mm}
\begin{tabular}{rlp{11.5cm}}
- & Exercise 2-1 : &  Transferring the (x,y,z) pixel cluster position to its $\eta$ and $\phi$ cylindrical coordinates and plotting the corresponding $\eta$ and $\phi$ distributions only in the case of the first barrel pixel layer (figures \ref{fig:ex2} (a) and (b)). \\
- & Exercise 2-2 : & Plotting the $\Delta\eta$ and $\Delta\phi$ distributions between the pixel cluster in the first barrel layer and the generator-level electron (figures \ref{fig:ex2} (c) and (d)). \\
- & Exercise 2-3 : & Plotting the $\Delta\eta$ and $\Delta\phi$ distributions between the pixel cluster in the first barrel layer and the L1 EM cluster (figures \ref{fig:ex2} (e) and (f)).  \\
\end{tabular}
\end{itemize}

\begin{figure}[tbp] 
\centering
  \begin{tabular}[t]{ccc}
\includegraphics[width=0.31\textwidth]{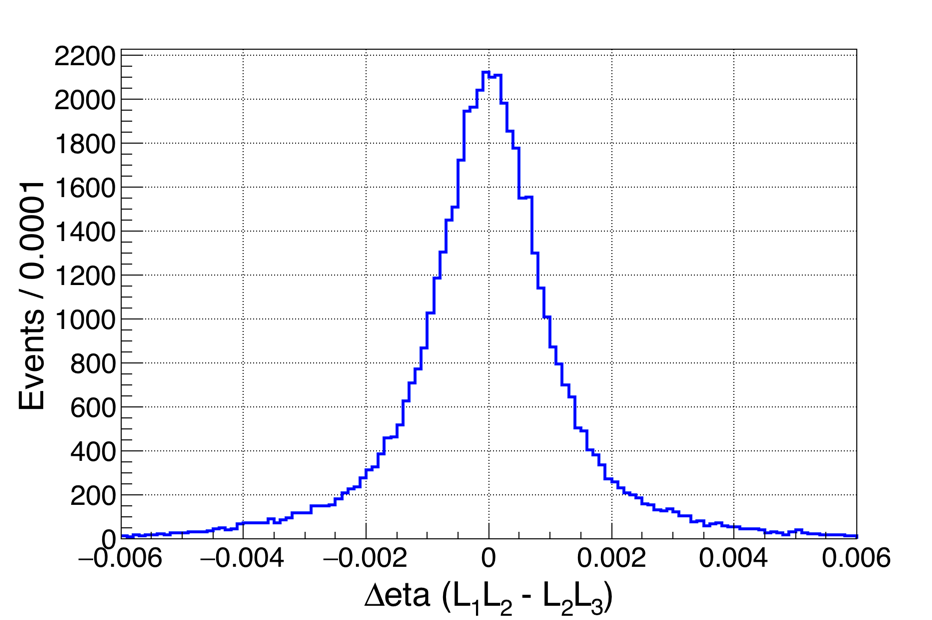} &
\includegraphics[width=0.31\textwidth]{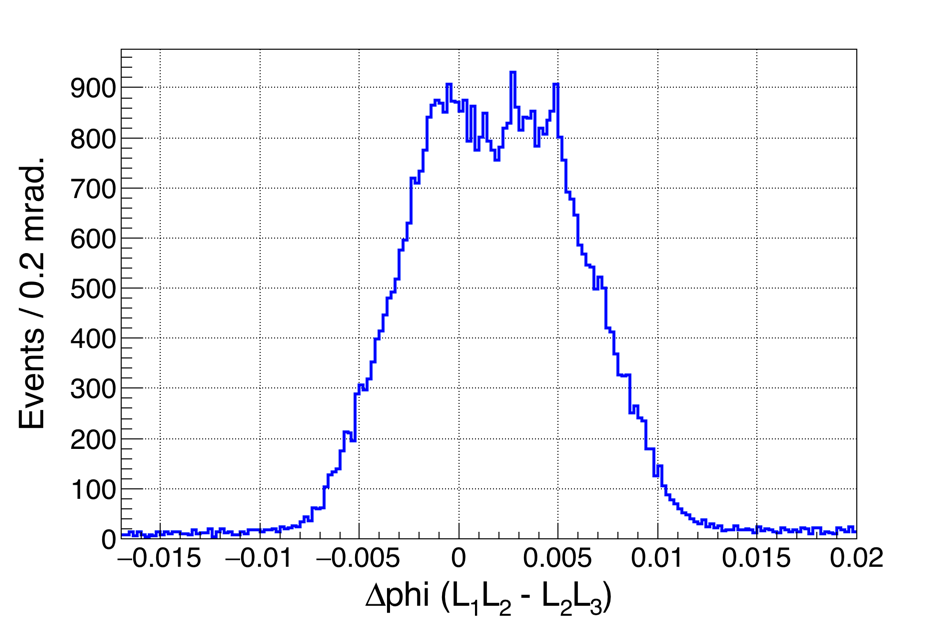} &
\includegraphics[width=0.31\textwidth]{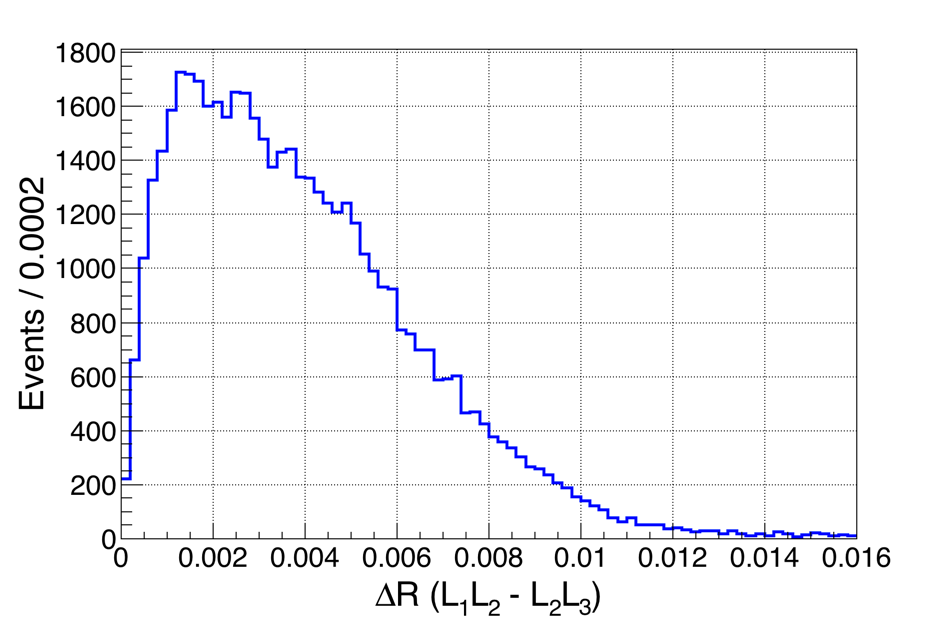} \\
(a) & (b) & (c) \\
\includegraphics[width=0.31\textwidth]{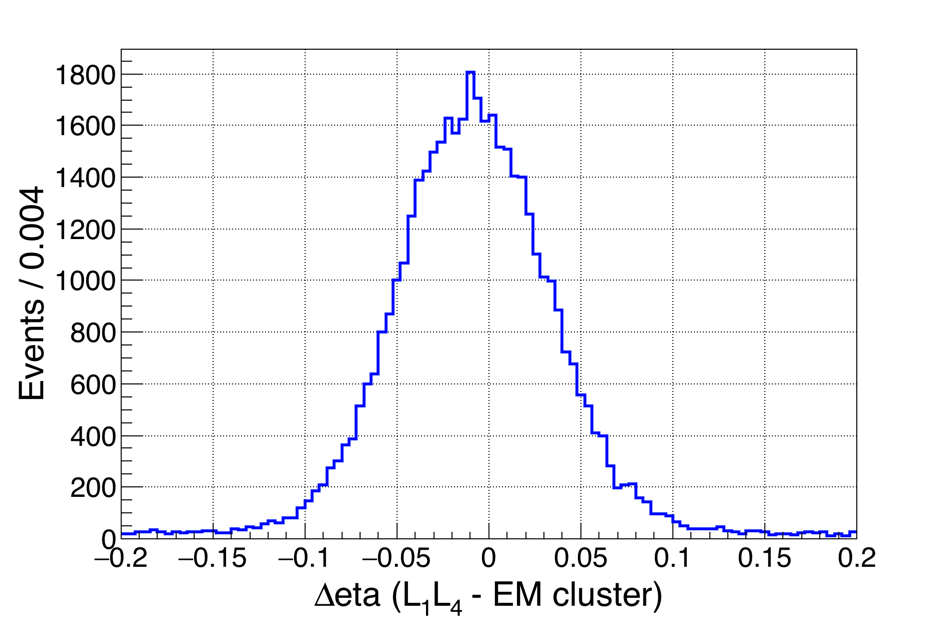} &
\includegraphics[width=0.31\textwidth]{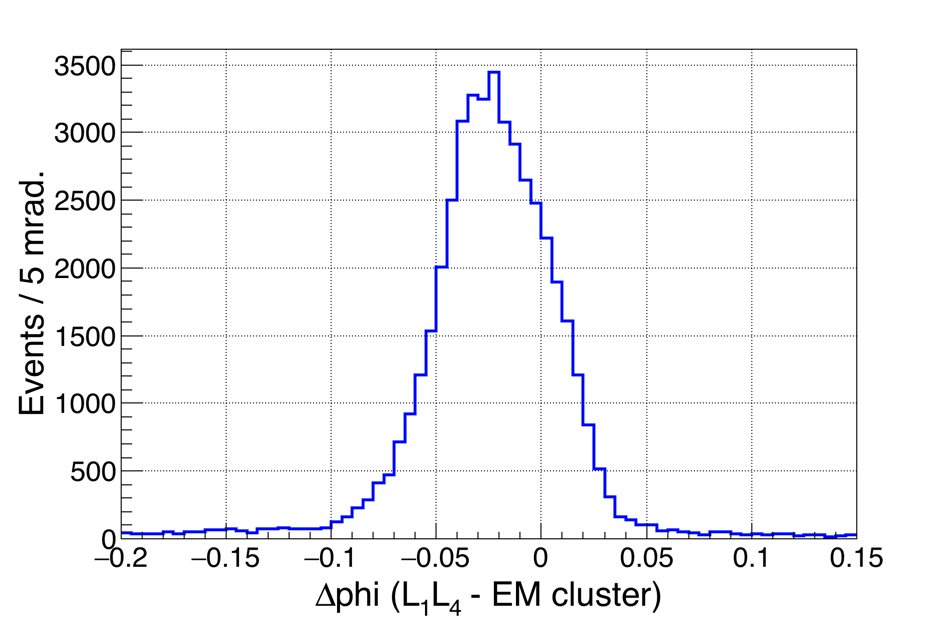} &
\includegraphics[width=0.31\textwidth]{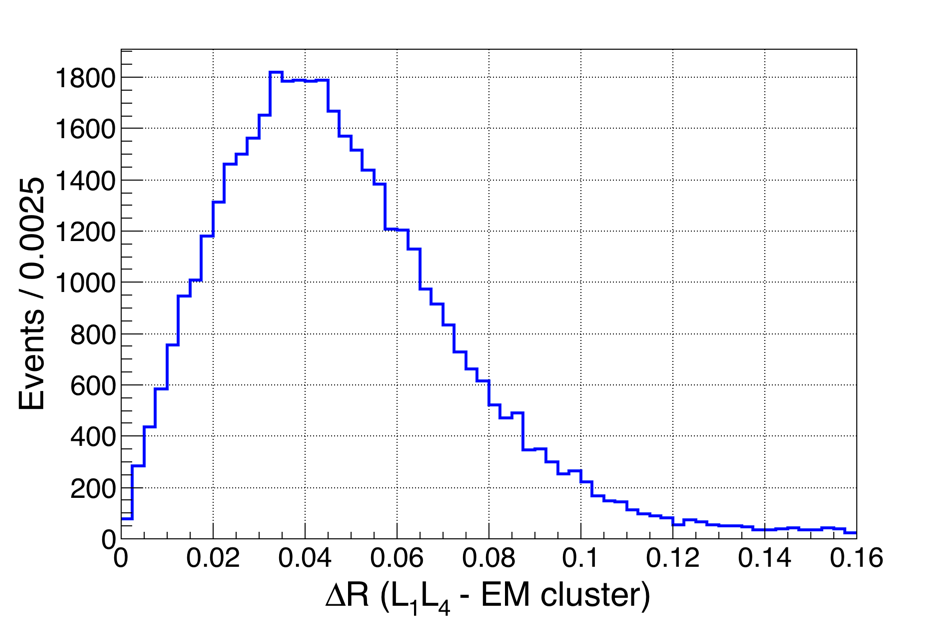} \\
(d) & (e) & (f) \\
  \end{tabular}
\caption{EXERCISE 3 results: Exercise 3-1: $\Delta\eta$ (a), $\Delta\phi$ (b) and $\Delta R$ (c) distributions between aligned clusters in the first and second layers and aligned clusters in the second and third layers;
Exercise 3-2: $\Delta\eta$ (d), $\Delta\phi$ (e) and $\Delta R$ (f) distributions
between colinear pixel clusters in the first and fourth layers and L1 EM cluster}
\label{fig:ex3}
\end{figure}

\begin{itemize}
\raggedright
  \item[(3)] The goal of the third exercise is to evaluate the size of the signal windows in two cases, one based on track segments using only the pixel clusters (see figure \ref{fig:standpixtrk}) i.e. {\it standalone pixel tracking}, the other one using tracks based on pixel clusters matching with the corresponding L1 EM cluster as shown for instance in figures \ref{fig:pixEM1} and \ref{fig:pixEM2} (b) i.e. {\it pixel track seeded by EM cluster}.
  
The first case (pixel standalone tracking) of the third exercise establishes the $\Delta\eta$, $\Delta\phi$ and $\Delta R$ signal windows using the corresponding pixel clusters in the first, second and third pixel layers. To do so, the students have simply to edit the basic generic source code provided to them. This source code allows extracting from the Ntuple the needed parameters. In this case it is used to compare  $\Delta\eta$, $\Delta\phi$ and $\Delta R$ angle differences between the track segment defined with the corresponding aligned clusters in the first and second layers and the track segment defined with the corresponding aligned clusters in the second and third layers for determining the corresponding $\eta$ and $\phi$ signal windows (figure \ref{fig:standpixtrk} (b)).

The second case (EM cluster seed case) involves deriving the distributions of $\Delta\eta$ (figure \ref{fig:ex3} (d)), $\Delta\phi$ (figure \ref{fig:ex3} (e)) and $\Delta R$ (figure \ref{fig:ex3} (f)) distributions which are defined as the corresponding angular differences between the pixel track segments using aligned clusters in layers (1,4), and the segment linking the origin (0,0,0) to the L1 EM cluster.

\renewcommand{\arraystretch}{1.3}
\renewcommand{\tabcolsep}{1.0mm}
\begin{tabular}{rlp{11.5cm}}
- & Exercise 3-1 : & Plotting the $\Delta\eta$, $\Delta\phi$ and $\Delta R$ signal windows in the case of the standalone pixel tracking (figure \ref{fig:ex3} (a), (b) and (c)). \\
- & Exercise 3-2 : & Plotting the $\Delta\eta$, $\Delta\phi$ and $\Delta R$ signal windows in the case of the pixel track seeded by EM cluster (figure \ref{fig:ex3} (d), (e) and (f)). \\
\end{tabular}
\end{itemize}

\begin{itemize}
\raggedright
    \item[(4)] The goal of the fourth exercise is to calculate the signal efficiency and the rate reduction factor for L1 EM calorimeter trigger. This is based on determining all the signal windows. It means the signal windows defined both in the standalone case and the EM cluster seed case. 
In the standalone case, the signal windows are computed with three pixel clusters using four pixel layers (including the beam spot). In the EM cluster seed case, the signal windows are computed with two pixel clusters and their matching with the corresponding EM cluster for all the possible combinations using the four pixel layers.

The electron efficiency of this trigger is defined as the ratio of the total number of events satisfying all the requirements on all the calculated signal windows to the total number of events selected by the L1 EM calorimeter trigger and matching in $\Delta R <$ 0.2 with the generator-level electron (ensuring the L1 EM objects are true electrons).

The potential of an electron trigger is also determined by its capability to reject ``fake electron''. 
The ``fake electron'' are objects which pass the L1 EM trigger conditions although they are not true electrons.
Minimum bias data are used to study the fake electron contamination, noting however that this sample also contains real electrons. We apply to the minimum bias data sample the L1 EM trigger conditions in order to distinguish the fake electrons from the real ones.

With this exercise, the students get to know how to distinguish a real electron candidate from a fake electron.


The L1 electron trigger reduction factor is defined as the ratio of the total number of minimum bias events that satisfy all requirements on the signal windows to the total number of events selected by the L1 EM calorimeter trigger.

In the final step, the students learn how to calculate the trigger efficiency and the rate reduction factor for L1 electron trigger as a function of the L1 EM $E_T$ trigger threshold.

 \renewcommand{\arraystretch}{1.3}
\renewcommand{\tabcolsep}{1.0mm}
\begin{tabular}{rlp{11.5cm}}
- & Exercise 4-1 : & Plotting the electron trigger efficiency as a function of L1 EM $E_T$ trigger threshold (figure \ref{fig:ex4} (a))\\
- & Exercise 4-2 : & Plotting the L1 electron trigger rate reduction factor as a function of L1 EM $E_T$ trigger threshold (figure \ref{fig:ex4} (b))\\
\end{tabular}
\end{itemize}

\begin{figure}[tbp] 
\centering
  \begin{tabular}[t]{cc}
\includegraphics[width=0.45\textwidth]{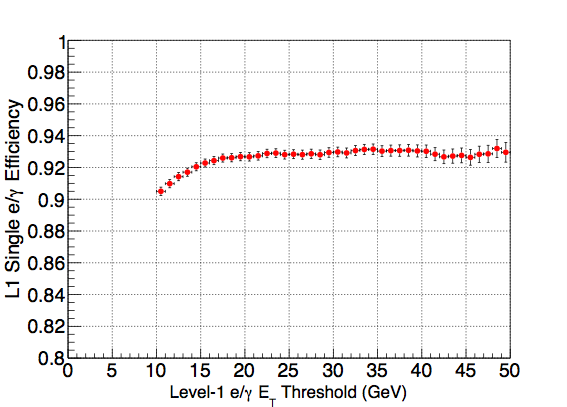} &
\includegraphics[width=0.45\textwidth]{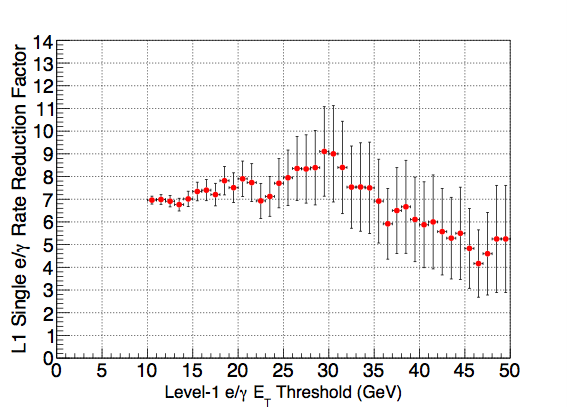} \\
(a) & (b) \\
  \end{tabular}
\caption{EXERCISE 4 results: Exercise 4-1: pixel based trigger efficiency (a) and Exercise 4-2: L1 electron trigger rate reduction factor with pixel information (b) as a function of L1 EM $E_T$ threshold}
\label{fig:ex4}
\end{figure}

\section{Conclusion}\label{sec:conclu}

This series of exercises aims to introduce the students to an important topic for future HEP experiments namely the potential use of the pixel detector in the L1 trigger or the real-time tracking.
It also shows the basic way to address the real-time tracking based on the pixel hits by building track segments in standalone or seed based tracking modes.
This is indeed a very challenging and new tracking trigger approach especially in the harsh HL-LHC environment. The main issues to be confronted are the bandwidth and trigger latency. This implies a lot of R\&D work on the hardware and software sides (new pixel front-end readout ASIC, new data processing architecture, advanced processor technology and associated software for high performance fast computing, new real-time algorithms development). Such R\&D efforts have started in these various aspects both in ATLAS and CMS. They are completed by feasibility studies that stress the benefits of such a trigger.

\acknowledgments

We acknowledge the support from the EU community Marie Curie International Incoming Fellowship (IIF), 
FP7-PEOPLE-2011-IIF, Contract No. 302103, TauKitforNewPhysics (Tau Toolkit for opening the New Physics Window at LHC and possible spin off effects).
And we thank our colleagues from the CMS collaboration for providing the overall CMS simulation framework.

\end{document}